\documentclass[aps,prl,notitlepage,twocolumn,superscriptaddress]{revtex4-1}

\usepackage{amsmath}
\usepackage{amssymb}
\usepackage{graphicx}
\usepackage{epstopdf}
\usepackage[colorlinks=true]{hyperref}

\newcommand{\tr}{\mathrm{T}}
\DeclareMathOperator{\sgn}{sgn}

\begin{document}
\title{Tunable Magnonic Thermal Hall Effect in Skyrmion Crystal Phases of Ferrimagnets}

\author{Se Kwon Kim}
\affiliation{Department of Physics and Astronomy, University of California, Los Angeles, California 90095, USA}
\affiliation{Department of Physics and Astronomy, University of Missouri, Columbia, Missouri 65211, USA}
\author{Kouki Nakata}
\affiliation{Advanced Science Research Center, Japan Atomic Energy Agency, Tokai, Ibaraki 319-1195, Japan}
\author{Daniel Loss}
\affiliation{Department of Physics, University of Basel, Klingelbergstrasse 82, CH-4056 Basel, Switzerland}
\affiliation{RIKEN Center for Emergent Matter Science (CEMS), Wako, Saitama 351-0198, Japan}
\author{Yaroslav Tserkovnyak}
\affiliation{Department of Physics and Astronomy, University of California, Los Angeles, California 90095, USA}

\begin{abstract}
We theoretically study the thermal Hall effect by magnons in skyrmion crystal phases of ferrimagnets in the vicinity of the angular momentum compensation point (CP). To this end, we start by deriving the equation of motion for magnons in the background of an arbitrary equilibrium spin texture, which gives rise to the fictitious electromagnetic field for magnons. As the net spin density varies, the resultant equation of motion interpolates between the relativistic Klein-Gordon equation at CP and the nonrelativistic Schr{\"o}dinger-like equation away from it. In skyrmion crystal phases, the right- and the left-circularly polarized magnons with respect to the order parameter are shown to form the Landau levels separately within the uniform skyrmion-density approximation. For an experimental proposal, we predict that the magnonic thermal Hall conductivity changes its sign when the ferrimagnet is tuned across CP, providing a way to control heat flux in spin-caloritronic devices on the one hand and a feasible way to detect CP of ferrimagnets on the other hand.
\end{abstract}

\date{\today}
\maketitle

\emph{Introduction.}|Magnetic skyrmions are swirling spin textures, which are characterized by the topological skyrmion number defined in terms of the real-space spin configuration~\cite{BogdanovJETP1989, *BogdanovJMMM1994}. Their topological characteristic influences not only the dynamics of themselves, e.g., by engendering the Magnus force, but also the dynamics of electrons moving through them~\cite{*[][{, and references therein.}] NagaosaNN2013}. In particular, when ferromagnetic skyrmions form a crystal lattice, electrons whose spin follows the local spin texture adiabatically experience the Lorentz force due to the fictitious magnetic field proportional to the skyrmion density and thereby exhibit the so-called topological Hall effect~\cite{LeePRL2009, *NeubauerPRL2009}. Recently, there has been a growing interest in skyrmion crystals in antiferromagnets and their electronic transport properties~\cite{GomonayNP2018, *SmejkalNP2018, BuhlPSS2017, *GobelPRB2017, *AkosaarXiv2017} because of their fundamental differences from ferromagnetic counterparts as well as technological applications for THz-speed magnetic devices~\cite{*[][{, and references therein.}] JungwirthNN2016}.

Magnons, which are quanta of spin waves~\cite{BlochZP1930, *HolsteinPR1940}, can transport information and exhibit topological phenomena similarly to electrons. Their potential ability to realize devices based on insulating magnets, which are free from drawbacks of conventional electronics such as significant energy loss due to Ohmic heating, has led to the emergence of magnon-based spintronics~\cite{*[][{, and references therein.}] ChumakNP2015}. In skyrmion crystal phases of ferromagnets, magnons have been shown to experience the fictitious magnetic field by keeping their spin antiparallel to the local spin texture~\cite{ShekaPRB2004, *DugaevPRB2005}. As a result, magnons form the approximate Landau levels with the finite Berry curvature~\cite{*[][{, and references therein.}] GarstJPD2017}, causing the thermal Hall effect~\cite{vanHoogdalemPRB2013, NakataPRB2017, SchuttePRB2014-2, *OhPRB2015, *OnoseScience2010, *ShiomiPRB2017}. However, the magnon bands and their transport properties in antiferromagnetic skyrmion crystals have not been studied.

\begin{figure}
\includegraphics[width=0.7\columnwidth]{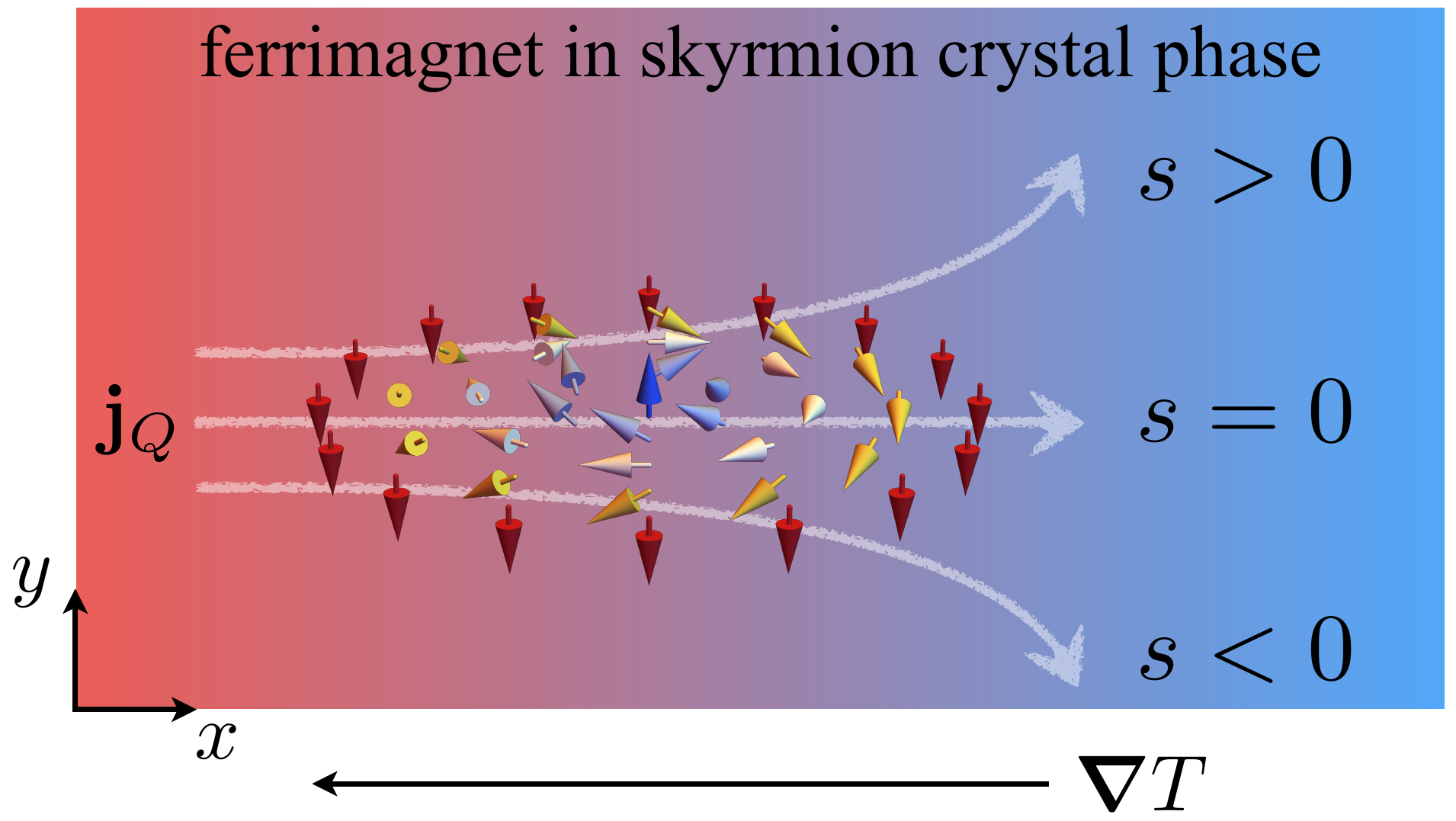}
\caption{Schematic illustration of the magnonic heat flux $\mathbf{j}_Q$ through a ferrimagnet in its skyrmion crystal phase subjected to a temperature gradient $\boldsymbol{\nabla} T$. The colored small arrows depict a single skyrmion texture of the order parameter $\mathbf{n}$. Magnons can exhibit the thermal Hall effect since the skyrmion crystal gives rise to the fictitious magnetic field, which magnons of left and right circular polarization (with respect to the order parameter) experience as if they carry the positive and the negative charge, respectively. The induced transverse heat flux changes its direction as the net spin density $s$ along $\mathbf{n}$ varies across $0$.}
\label{fig:fig1}
\end{figure}

In this Letter, we fill this gap by investigating a theoretically more general problem: The dynamics of magnons in the presence of skyrmion crystals in ferrimagnets exhibiting the angular momentum compensation point (CP)~\cite{*[][{, and references therein.}] KirilyukRPP2013}, at which the net spin density vanishes, but the magnetization can be finite. One class of such ferrimagnets is rare-earth transition-metal alloys such as GdFeCo or CoGd whose net spin density can be tuned by varying either temperature~\cite{StanciuPRB2006} or chemical composition~\cite{BinderPRB2006}. To this end, we start by deriving the equation of motion for magnons in the presence of an arbitrary spin texture, which includes the fictitious electromagnetic field. The obtained equation of motion is reduced to the nonrelativistic Schr{\"o}dinger-like equation for ferromagnetic magnons away from CP and to the relativistic Klein-Gordon equation for antiferromagnetic magnons at CP, interpolating between the dynamics of ferromagnets and that of antiferromagnets as previously shown for the dynamics of domain walls and skyrmions~\cite{KimNM2017, KimPRB2017, *OhPRB2017}. In the presence of a skyrmion crystal, two species of magnons with right and left circular polarization (with respect to the order parameter) will be shown to experience the fictitious magnetic fields of opposite directions and form the Landau levels separately, realizing two-dimensional magnonic topological insulators~\cite{NakataPRB2017-2}. As an experimental proposal, we will show that the thermal Hall conductivity changes its sign as the ferrimagnet is tuned across CP. See Fig.~\ref{fig:fig1} for a schematic illustration. One promising platform is offered by GdFeCo films, where isolated skyrmions have been observed~\cite{WooNC2018, *WooarXiv2017} and the antiferromagnetic domain-wall dynamics has been demonstrated at CP~\cite{KimNM2017}. The proposal provides not only a feasible way to control the direction of the thermal flux, which can be useful in spin caloritronics~\cite{BauerNM2012}, but also a thermal transport measurement for determining CP, which can complement the other methods based on magnetic resonances~\cite{StanciuPRB2006, KimAPL2017} or domain-wall speed measurements~\cite{KimNM2017}.

\emph{General formalism.}|Our model system is a collinear ferrimagnet, whose potential energy density is given by
\begin{equation}
\mathcal{U}[\mathbf{n}] = A (\boldsymbol{\nabla} \mathbf{n})^2 / 2 + D \mathbf{n} \cdot (\boldsymbol{\nabla} \times \mathbf{n}) - \mathbf{h} \cdot \mathbf{n} \, ,
\end{equation}
where $\mathbf{n} (\mathbf{r}, t)$ is the three-dimensional unit vector representing the direction of the magnetic order. Here, the first term is the exchange energy; the second term is the Dzyaloshinskii-Moriya interaction (DMI), which exists when the inversion symmetry is broken~\footnote{Although we considered only one type of spin-orbit coupling, $\propto \mathbf{n} \cdot (\boldsymbol{\nabla} \times \mathbf{n})$, in this work, our main results of the magnonic Landau levels and the ensuing thermal Hall effect are valid as long as the ferrimagnet is in the skyrmion crystal phase regardless of the type of spin-orbit coupling.}; the last term represents the Zeeman coupling between the external field $\mathbf{h}$ and the magnetization along the direction of the order parameter. Here, we are neglecting the other terms such as the dipolar interaction by following the previous literature on chiral magnets~\cite{HanPRB2010, *PetrovaPRB2011, *BanerjeePRX2014}. With a suitable choice of the coefficient values, the ground state is a skyrmion crystal~\cite{HanPRB2010}, which will be the phase of our main interest for later discussions. The equilibrium order-parameter configuration will be denoted by $\mathbf{n}_0$.

The dynamics of the order parameter $\mathbf{n}$ of the ferrimagnet can be described by the following Landau-Lifshitz-like equation~\cite{IvanovSSC1984, *LossPRL1992, KimPRB2017, KimNM2017}:
\begin{equation}
\label{eq:llg}
s \dot{\mathbf{n}} + \rho \mathbf{n} \times \ddot{\mathbf{n}} = \mathbf{n} \times \left(A \nabla^2 \mathbf{n} - 2 D \boldsymbol{\nabla} \times \mathbf{n} + \mathbf{h} \right) \, ,
\end{equation}
where $s$ is the equilibrium spin density and $\rho$ parametrizes the inertia associated with the dynamics of the order parameter. The left-hand side is the time derivative of the net spin density, $\mathbf{s} = s \mathbf{n} + \rho \mathbf{n} \times \dot{\mathbf{n}}$, the former and the latter of which are the longitudinal and the transverse component of the spin density with respect to the order parameter, respectively~\cite{AndreevSPU1980}. Conventional ferromagnets and antiferromagnets have only the first and the second term, respectively, on the left-hand side. The parameter of focus in this work is the spin density $s$, which can be varied across zero. We are interested in the change of the magnon bands as a function of it.

To obtain the equation of motion for a magnon, which is a quantum of small-amplitude fluctuations from the equilibrium state, we use the local coordinate system $\mathbf{n}'$, where the equilibrium state is in the positive $z$ direction, $\mathbf{n}'_0 \equiv \hat{\mathbf{z}}$~\cite{KovalevEPL2012, *KovalevPRB2014, vanHoogdalemPRB2013}. The transformation can be implemented by a three dimensional rotation matrix $\mathcal{R}$ satisfying $\mathbf{n}_0 = \mathcal{R} \mathbf{n}'_0$. We will use one explicit realization of it in this work: $\mathcal{R} = \exp(\phi_0 \mathcal{L}_z) \exp(\theta_0 \mathcal{L}_y)$ for $\mathbf{n}_0 = (\sin \theta_0 \cos \phi_0, \sin \theta_0 \sin \phi_0, \cos \theta_0)$ where $\mathcal{L}_y$ and $\mathcal{L}_z$ are the generators of the rotations about the $y$ and the $z$ axis, respectively~\footnote{The generators for three-dimensional rotations can be defined in terms of the Levi-Civita symbol as follows: $(\mathcal{L}_a)_{bc} = - \epsilon_{abc}$, where $a, b,$ and $c$ are indices for spatial coordinates, $x, y,$ and $z$.}. 

The equation of motion for magnons can be obtained from Eq.~(\ref{eq:llg}) to linear order in the fluctuation $\mathbf{\delta n}' \equiv \mathbf{n}' - \hat{\mathbf{z}} = n'_x \hat{\mathbf{x}} + n'_y \hat{\mathbf{y}}$. Since the equation is second order in time derivative, there are two types of solutions. It is convenient to represent the two monochromatic solutions with the complex fields: $\psi_+ =  n'_x - i n'_y \propto \exp(- i \epsilon t / \hbar)$ for right-circularly polarized magnons and $\psi_- = n'_x + i n'_y \propto \exp(- i \epsilon t / \hbar)$ for left-circularly polarized magnons where $\epsilon$ is the magnon energy. We will refer the former and the latter to the positive-chirality ($q = 1$) and the negative-chirality ($q = -1$) solutions, respectively. The equation of motion for a magnon of chirality $q$ is given by
\begin{equation}
\label{eq:eom}
- q s \left( i \partial_t - \frac{q \phi}{\hbar} \right) \psi_q + \rho \left( i \partial_t - \frac{q \phi}{\hbar} \right)^2 \psi_q = A \left( \frac{\boldsymbol{\nabla}}{i} - \frac{q \mathbf{a}}{\hbar} \right)^2 \psi_q \, ,
\end{equation}
which is our first main result. See Supplemental Material for its derivation~\footnote{Supplemental Material contains the derivation of the equations of motion for magnons and the summary of the known results for the Landau levels of a charged particle.}. Here, $\phi$ is the texture-induced scalar potential given by $\phi = \hbar (\mathcal{R}^{-1} \partial_t \mathcal{R})_{12} = - \hbar \cos \theta_0 \partial_t \phi_0${, where $\mathcal{M}_{12}$ represents a corresponding matrix element of $\mathcal{M}$. The vector potential consists of two contributions~\cite{vanHoogdalemPRB2013}: $\mathbf{a} = \mathbf{a}^t + \mathbf{a}^d$, where the first term is from the exchange energy, $a^t_i = -\hbar (\mathcal{R}^{-1} \partial_i \mathcal{R})_{12} = \hbar \cos \theta_0 \partial_i \phi_0$, and the second term is from the DMI, $\mathbf{a}^d = - (\hbar D/A) \mathbf{n}_0$. The texture-induced fictitious electric and magnetic fields are given by
\begin{eqnarray}
e^t_i &=& - \partial_i \phi - \partial_t a^t_i = \hbar \mathbf{n}_0 \cdot (\partial_t \mathbf{n}_0 \times \partial_i \mathbf{n}_0) \, , \\
b^t_i &=& \epsilon_{ijk} \partial_j a^t_k = (\hbar / 2) \epsilon_{ijk} \mathbf{n}_0 \cdot (\partial_k \mathbf{n}_0 \times \partial_j \mathbf{n}_0) \, , \label{eq:b}
\end{eqnarray}
in the Einstein summation convention. The obtained fields are identical to those for electrons~\cite{VolovikJPC1987, *WongPRB2009, *ZangPRL2011, NagaosaNN2013} and magnons~\cite{KovalevEPL2012, vanHoogdalemPRB2013} in ferromagnets. The DMI-induced vector potential $\mathbf{a}^d$ gives rise to another contribution to the fictitious fields, $\mathbf{e}^d = - \partial_t \mathbf{a}^d = (\hbar D / A) \partial_t \mathbf{n}_0$ and $\mathbf{b}^d = - (\hbar D / A) \boldsymbol{\nabla} \times \mathbf{n}_0$~\cite{vanHoogdalemPRB2013}. Note that the chirality $q = \pm 1$ of a magnon serves as its charge with respect to the electromagnetic fields, which can be understood as follows. Since the positive- and negative-chirality magnons carry spin whose directions are locked parallel and antiparallel with respect to the background spin texture $\mathbf{n}_0$~\cite{TvetenPRL2014}, they pick up the Berry phase with the opposite signs and experience the opposite fictitious electromagnetic fields~\cite{ChengPRB2012}. During the derivation, we neglected the second and higher order terms in $\phi$ and $\mathbf{a}$ and the term proportional to the external field, by focusing on high-energy magnons whose wavelength is much smaller than the spatial extension of the texture and whose kinetic energy dominates the Zeeman energy, which we will refer to as the exchange approximation~\footnote{Here, we remark that all the potential terms, $\phi, \mathbf{a}^t$, and $\mathbf{a}^d$, are functions of the material parameters such as $A, D$, and $\mathbf{h}$.}.

At CP, the equilibrium spin density vanishes, $s = 0$, and the nature of the dynamics becomes antiferromagnetic. The equation of motion is then reduced to the following Klein-Gordon equation~\cite{SchrodingerAP1926, *GordonZP1926, *KleinZP1927} which describes the dynamics of a relativistic particle with charge $q$ in the presence of an electromagnetic field:
\begin{equation}
\label{eq:afm-eom}
\left( i \hbar \partial_t - q \phi \right)^2 \psi_q = c^2  \left( \frac{\hbar}{i} \boldsymbol{\nabla} - q \mathbf{a} \right)^2 \psi_q \, ,
\end{equation}
where $c \equiv \sqrt{A / \rho}$ is the characteristic speed that is the magnon speed in the absence of electromagnetic fields. This equation describing the dynamics of magnons in antiferromagnets moving through a general spin texture has not been derived before except for a special case of a one-dimensional domain wall~\cite{TvetenPRL2014, *KimPRB2014, *QaiumzadehPRB2018}. 

When sufficiently distant from CP, a ferrimagnet has enough spin density $s$ to neglect the inertial term $\propto \rho$ in Eq.~(\ref{eq:llg}) for the low-energy dynamics. The equation of motion~(\ref{eq:eom}) for ferrimagnetic magnons is then reduced to that for ferromagnetic magnons~\cite{KovalevEPL2012, vanHoogdalemPRB2013}:
\begin{equation}
\label{eq:fm-eom}
- \sgn(q s) i \hbar \partial_t \psi_q = \left[ \frac{1}{2 m} \left( \frac{\hbar}{i} \boldsymbol{\nabla} - q \mathbf{a} \right)^2 - \sgn(s) \phi \right] \psi_q \, ,
\end{equation}
with the effective mass $m = \hbar |s| / 2 A$, which resembles the Schr{\"o}dinger equation for a nonrelativistic charged particle subjected to an electromagnetic field. 

It is instructive to discuss the solutions to Eq.~(\ref{eq:eom}) in the absence of the scalar and the vector potentials $\phi = 0$ and $\mathbf{a} = \mathbf{0}$, which are given by the plane-wave solutions, $\psi_q \propto \exp(i \mathbf{k} \cdot \mathbf{r} - i \epsilon t / \hbar)$~\cite{ChengSR2016}. The energy-momentum relation is given by 
\begin{equation}
\epsilon_q (\mathbf{k}) = \sqrt{(m c^2)^2 + \hbar^2 c^2 \mathbf{k}^2} + \sgn(q s) m c^2 \, .
\end{equation}
The solution to Eq.~(\ref{eq:afm-eom}) for an antiferromagnetic magnon is given by the high-kinetic energy limit, i.e., $\hbar |\mathbf{k}| \gg m c$: $\epsilon_\pm \approx \hbar c |\mathbf{k}|$. The two solutions are degenerate at the level of approximation taken in Eq.~(\ref{eq:afm-eom}) where the time reversal symmetry is respected by having vanishing spin density $s = 0$. The lower-energy solution to Eq.~(\ref{eq:fm-eom}) for a ferromagnetic magnon is given by the low-kinetic energy limit, i.e., $\hbar |\mathbf{k}| \ll m c$: $\epsilon_q \approx \hbar^2 \mathbf{k}^2 / 2 m$ with the chirality $q = - \sgn(s)$. Note that spin of the low-energy magnons is locked antiparallel to the direction of the background spin density $s \mathbf{n}_0$. Here, the momentum scale governing the separation between a nonrelativistic and a relativistic regime is given by $m c = \hbar |s| / 2 \sqrt{A \rho}$.

\emph{Magnon in a skyrmion crystal.}|Now, let us apply the above formalism to one specific example: Magnons in a skyrmion crystal of a quasi-two-dimensional ferrimagnet. We will assume that the skyrmion crystal is static, for which the fictitious electric field vanishes. A skyrmion is characterized by its integer skyrmion number~\cite{BelavinJETP1975}:
\begin{equation}
Q = \frac{1}{4 \pi} \int dx dy \, \mathbf{n}_0 \cdot (\partial_x \mathbf{n}_0 \times \partial_y \mathbf{n}_0) \equiv \int dxdy \, \rho_{sky} \, ,
\end{equation}
counting how many times the order parameter $\mathbf{n}_0$ wraps the unit sphere. Under suitable conditions, skyrmions with the definite skyrmion number can crystalize in the triangular lattice as observed in several ferromagnetic materials~\cite{NagaosaNN2013}, giving rise to the finite skyrmion number density per unit area, which we denote by $\rho_{sky}$. The associated fictitious magnetic field [Eq.~(\ref{eq:b})] is given by 
\begin{equation}
\mathbf{b}^t = - 4 \pi \hbar \rho_{sky} \hat{\mathbf{z}} \, .
\end{equation}
The spatial profile of the magnetic field depends on the detailed values of material parameters, making it cumbersome to take into account analytically. Therefore, below, we will account for its effects by spatially averaging it: $\mathbf{b} = - 4 \pi \hbar \langle \rho_{sky} \rangle \hat{\mathbf{z}}$. The corresponding magnetic length is given by $l = \sqrt{\hbar / |b_z|} = 1 / \sqrt{4 \pi |\langle \rho_{sky} \rangle|}$, which is proportional to the distance between neighboring skyrmions. The DMI-induced contribution $\mathbf{b}^d$ vanishes after spatial averaging. In addition, we will assume the negative skyrmion density $\rho_{sky} < 0$ and thus $b_z > 0$ without loss of generality for subsequent discussions.
\begin{figure}
\includegraphics[width=0.8\columnwidth]{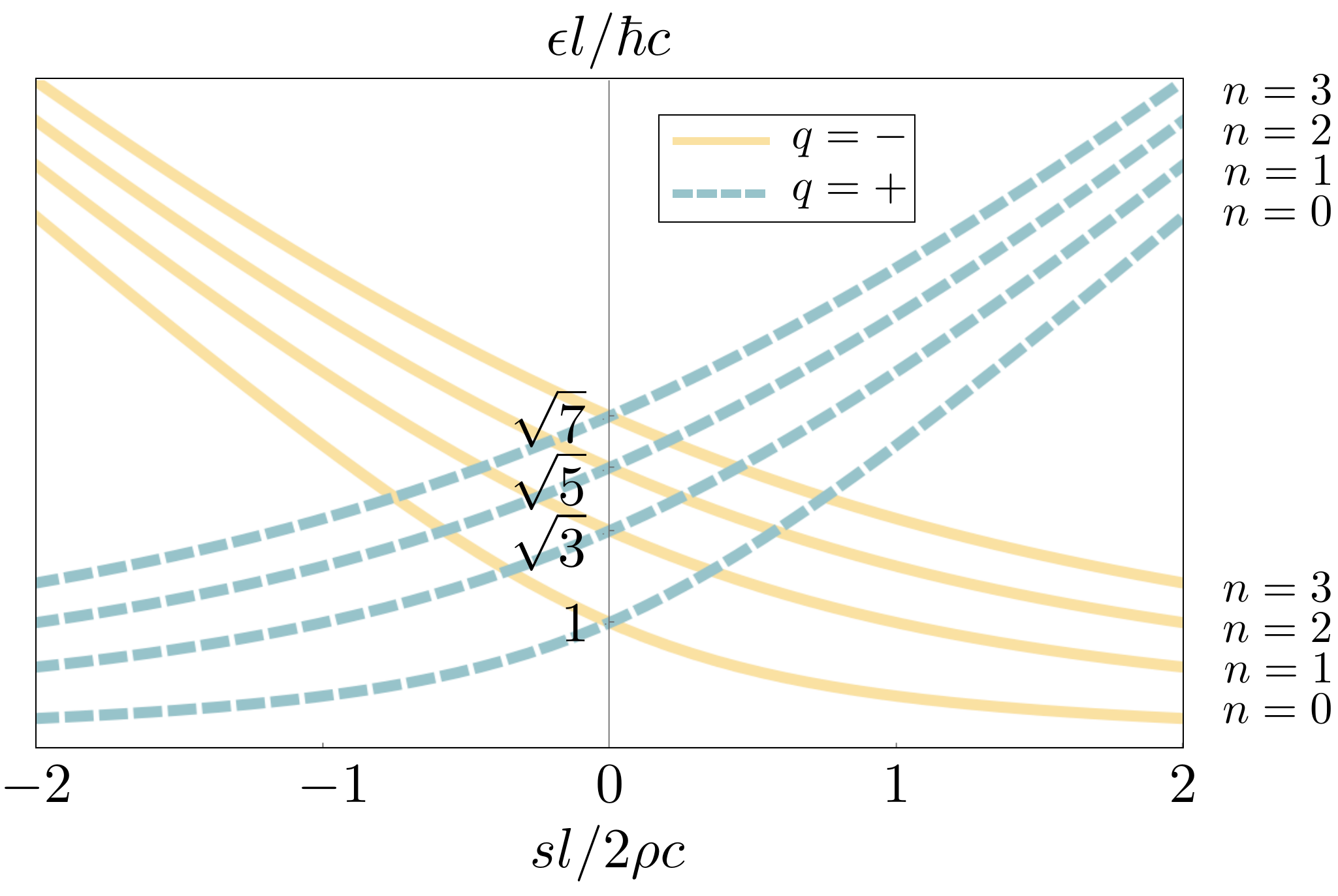}
\caption{The plot of the Landau levels [Eq.~(\ref{eq:soln})] of magnon bands in ferrimagnetic skyrmion crystals in terms of the rescaled energy $\epsilon l / \hbar c$ and the rescaled spin density $s l / 2 \rho c$. The solid gold and the dashed blue lines represent the right-circularly polarized ($q = +$) and the left-circularly polarized ($q = -$) magnon bands, respectively.}
\label{fig:fig2}
\end{figure}

To solve Eq.~(\ref{eq:eom}), we adopt the known results for the Landau levels of a nonrelativistic charged particle subjected to a uniform magnetic field~\cite{[][{, and references therein.}] Girvin1999, Note3}. Plugging the monochromatic function, $\psi_q (\mathbf{r}, t) = \exp(- i \epsilon t / \hbar) \psi_{n n'} (\mathbf{r})$ into Eq.~(\ref{eq:eom}), where $\psi_{n n'} (\mathbf{r})$ is the known eigenfunction of the right-hand side of Eq.~(\ref{eq:eom}) for the $n$th Landau level ($n'$ is the index for states within each Landau level), yields the following solutions:
\begin{equation}
\label{eq:soln}
\epsilon^q_n = \sqrt{(m c^2)^2 + \hbar c^2 b_z (2 n + 1)} + \sgn(q s) m c^2 \, .
\end{equation}
These magnon bands in a ferrimagnetic skyrmion crystal within the approximation of the uniform skyrmion density is our second main result. The number of states in one Landau level is given by the total number of the fictitious magnetic flux quanta through the plane, which is twice the total number of skyrmions in the system. The massless relativistic limit is given by $\epsilon^\pm_n \approx c \sqrt{\hbar b_z (2 n+1)}$, which agrees with the known result for the Klein-Gordon equation~\cite{LamJMP1971}. The lower band in the massive limit, $m c ^2 \gg c \sqrt{ \hbar b_z ( 2 n +1)}$, is reduced to the solution for nonrelativistic particles: $\epsilon_n \approx (\hbar b_z / 2m) (2 n +1)$. The solution can be cast into the dimensionless form in terms of the rescaled energy $\xi \equiv \epsilon l / \hbar c$ and the rescaled spin density $\zeta \equiv s l / 2 \rho c$: $\xi^\pm_n = \sqrt{\zeta^2 + 2 n + 1} \pm \zeta$. See Fig.~\ref{fig:fig2} for the plot. The solid gold and the dashed blue line represent the right-circularly polarized ($q = +$) and the left-circularly polarized ($q = -$) magnon bands, respectively. The $x$ axis of Fig.~\ref{fig:fig2} can be swept through the origin by varying the temperature across CP, the physical implication of which is discussed below.

\begin{figure}
\includegraphics[width=0.8\columnwidth]{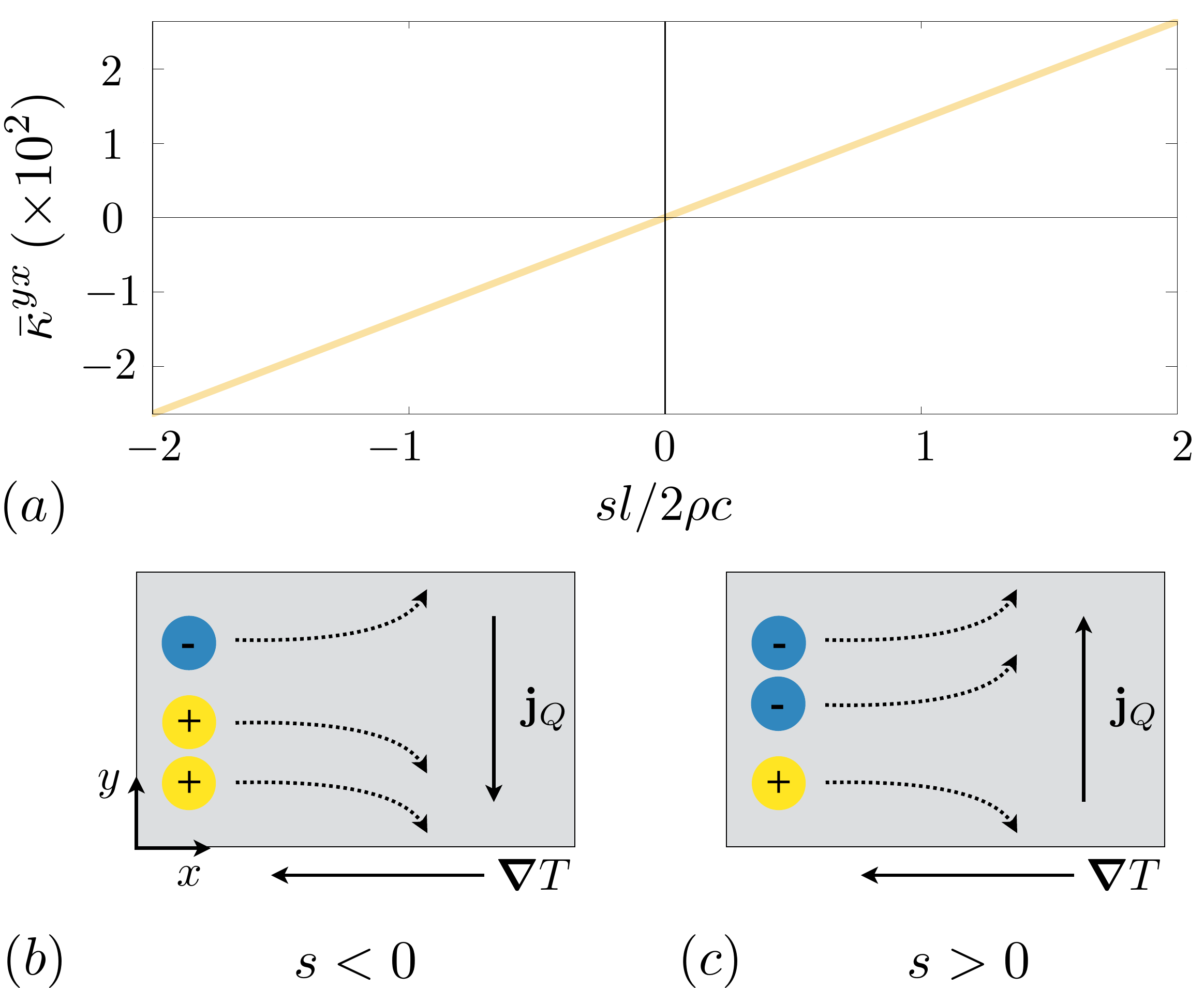}
\caption{(a) The plot of the rescaled thermal Hall conductivity $\bar{\kappa}^{yx} \equiv (2 \pi \hbar t / k_B^2 T) \kappa^{yx}$, which parametrizies the ratio of the induced transverse heat flux $j_Q^y$ to the applied longitudinal temperature gradient $\partial_x T$ for the ferrimagnet film of thickness $t$. (b) and (c) the schematic illustrations of the positive-chirality ($+$) and the negative-chirality ($-$) magnon motions subjected to a temperature gradient $\boldsymbol{\nabla} T = (\partial_x T) \hat{\mathbf{x}}$ and the resultant transverse energy flux $\mathbf{j}_Q = j_Q^y \hat{\mathbf{y}}$.}
\label{fig:fig3}
\end{figure}

\emph{Tunable thermal Hall effect.}|Each flat Landau level of magnons~\cite{vanHoogdalemPRB2013, NakataPRB2017} has the Chern number $\nu_0 = - q$, which is the integral of the uniform Berry curvature defined in terms of the magnonic wavefunction over the Brillouin zone~\cite{Note3}. Magnons with the finite Berry curvature can give rise to the thermal Hall effect~\cite{[][{, and references therein.}] MurakamiJPSJ2017, [][{, and references therein.}] XiaoRMP2010}, a phenomenon of the generation of the transverse energy flux $j_Q^y$ upon the application of the longitudinal temperature gradient $\partial_x T$: $j_Q^y = - \kappa^{yx} \partial_x T$, which is quantified by the thermal Hall conductivity $\kappa^{yx}$~\footnote{In this Letter, we do not include the effects of the nonequilibrium magnon accumulation on the thermal magnon transport, which can be important in a system with appropriate boundary conditions as shown in Ref.~\cite{NakataPRB2017}}. Within the linear response theory, the thermal Hall conductivity for our case is given by $\kappa^{yx} = (k_B^2 T / 2 \pi \hbar t) \sum_n [c_2 (\rho^B (\epsilon_n^-)) - c_2(\rho^B (\epsilon_n^+))]$, where $t$ is the thickness of the ferrimagnet, $c_2(x) = (1 + x) (\log(1 + x^{-1}))^2 - (\log x)^2 - 2 \text{Li}_2(- x)$, $\text{Li}_2 (z)$ is the polylogarithm function, and $\rho^B (\epsilon) = [\exp(\epsilon / k_B T) - 1]^{-1}$ is the Bose-Einstein distribution~\cite{MurakamiJPSJ2017}. Figure~\ref{fig:fig3}(a) shows the plot of the rescaled thermal Hall conductivity $\bar{\kappa}^{yx} \equiv (2 \pi \hbar t/ k_B^2 T) \kappa^{yx}$ as a function of the rescaled spin density $\zeta = s l / 2 \rho c$. The plot is drawn with the following value for the ratio of the characteristic relativistic energy scale to the temperature: $\hbar c l^{-1} / k_B T \sim 0.03$, which is obtained from the magnon speed $c \sim 10^4$ m/s calculated for GdFeCo~\cite{OhPRB2017}, the magnetic length $l = 1 / \sqrt{4 \pi |\langle \rho_{sky} \rangle|} \sim 10$ nm for the inter-skyrmion distance $\sim 50$ nm of triangular lattice of skyrmions observed in chiral ferromagnets~\cite{YuNature2010, *SekiScience2012}, and the temperature $T = 70$ K. The induced transverse heat flux changes its direction as the net spin density varies across zero, at which magnons of two chiralities are degenerate and thus the thermal Hall effect is absent similarly to antiferromagnets~\cite{NakataPRB2017-2}. When the spin density is negative, for example, there are more positive-chirality magnons than the others, causing the negative thermal Hall conductivity as shown in Fig.~\ref{fig:fig3}(b). Here, we remark that, although the numerical results shown in Fig.~\ref{fig:fig3}(a) are obtained within the approximation of the uniform fictitious magnetic field, the sign change of $\kappa^{yx}$ at $s = 0$ is the generic property of Eq.~(\ref{eq:eom}) under the time reversal and thus does not rely on the approximation.

Next, let us estimate the change of the thermal Hall conductivity $\Delta \kappa^{yx}$ as the net spin density $s$ varies by $\Delta s \sim 5 \times 10^{-8}$ J$\cdot$s/m$^3$, which can be achieved by changing the temperature by $\Delta T = 10$ K around CP of GdFeCo films according to the numerical results in Ref.~\cite{KimNM2017}. Here, we assume that all the parameters except the spin density are constant. By using the inertia $\rho \sim \hbar^2 / J d^3$ obtained with the Heisenberg exchange energy $J \sim 5$ meV and the lattice constant $d \sim 0.5$ nm, the above $\Delta s$ yields $\Delta \kappa^{yx} \sim 0.05$ W/K$\cdot$m for $50$-nm thick films, which is comparable to the large thermal Hall conductivities observed in frustrated magnets~\cite{HirschbergerScience2015, *HirschbergerPRL2015}.

\emph{Discussion.}|By investigating the magnon bands in a skyrmion crystal of a ferrimagnet in the vicinity of CP, we have shown that, under certain approximations, the positive-chirality and the negative-chirality magnons form Landau levels separately by experiencing the skyrmion-induced fictitious magnetic field, leading to the identification of ferrimagnets in their skyrmion-crystal phases as magnonic topological insulators. We have predicted that the sign of the resultant thermal Hall conductivity changes its sign across CP. We mention here that the Berry curvature of the magnon bands will cause the spin Nernst effect~\cite{BauerNM2012, KovalevPRB2016, *KimPRL2016-2, *ZyuzinPRL2016, *ChengPRL2016}, a phenomenon of the generation of a transverse spin current by a longitudinal temperature gradient, in addition to the thermal Hall effect. The former, respecting the time reversal symmetry, will be finite even at CP unlike the latter disappearing at CP.

\begin{acknowledgments}
S.K.K. acknowledges the enlightening discussions with Oleg Tchernyshyov and Ari Turner. S.K.K. and Y.T. were supported by the Army Research Office under Contract No. W911NF-14-1-0016. K.N. was supported by Leading Initiative for Excellent Young Researchers, MEXT, Japan. D.L. was supported by the Swiss National Science Foundation, the NCCR QSIT, and JSPS KAKENHI Grant Numbers 16K05411.
\end{acknowledgments}

\bibliography{/Users/evol/Dropbox/School/Research/master}

\cleardoublepage

\begin{widetext}
\setcounter{figure}{0}
\setcounter{equation}{0}
\renewcommand{\thefigure}{S\arabic{figure}}  
\renewcommand{\theequation}{S\arabic{equation}}
\begin{center}
{\bf Supplemental Material: Tunable Magnonic Thermal Hall Effect in Skyrmion Crystal Phases of Ferrimagnets}\vspace{0.2cm} \\
{Se Kwon Kim,$^{1,2}$ Kouki Nakata,$^3$ Daniel
  Loss,$^{4,5}$ and Yaroslav Tserkovnyak$^1$}\\
$^1${\small\it Department of Physics and Astronomy, University of California, Los Angeles, California 90095, USA} \\
$^2${\small\it Department of Physics and Astronomy, University of Missouri, Columbia, Missouri 65211, USA} \\
$^3${\small\it Advanced Science Research Center, Japan Atomic Energy Agency, Tokai, Ibaraki 319-1195, Japan} \\
$^4${\small\it Department of Physics, University of Basel, Klingelbergstrasse 82, CH-4056 Basel, Switzerland} \\
$^5${\small\it RIKEN Center for Emergent Matter Science (CEMS), Wako, Saitama 351-0198, Japan} \\
{\small(Dated: August 20, 2018)}\vspace{0.5cm}\\
\begin{minipage}[ht]{0.8\textwidth}
{\small\hspace{0.3cm}This Supplemental Material contains the derivation of the equations of motion for magnons and the summary of the known results for the Landau levels of a charged particle.}
\end{minipage}
\end{center}

\section{The derivation of the emergent electromagnetic fields.}

We use the three-dimensional rotation matrix $\mathcal{R}$ that transforms the $z$ axis to the equilibrium configuration $\mathbf{n}_0$: $\mathbf{n}_0 = \mathcal{R} \mathbf{n}'_0$ with $\mathbf{n}'_0 \equiv \hat{\mathbf{z}}$. One explicit form for $\mathcal{R}$ is given by $\mathcal{R} = \exp(\phi_0 \mathcal{L}_z) \exp(\theta_0 \mathcal{L}_y)$ for $\mathbf{n}_0 = (\sin \theta_0 \cos \phi_0, \sin \theta_0 \sin \phi_0, \cos \theta_0)$ where $\mathcal{L}_x, \mathcal{L}_y$, and $\mathcal{L}_z$ are the generators of the three-dimensional rotations about the $x, y,$ and $z$ axis given by
\begin{equation}
\mathcal{L}_x = \begin{pmatrix} 0 & 0 & 0 \\ 0 & 0 & -1 \\ 0 & 1 & 0 \end{pmatrix} \, , \quad \mathcal{L}_y = \begin{pmatrix} 0 & 0 & 1 \\ 0 & 0 & 0 \\ -1 & 0 & 0 \end{pmatrix} \, , \quad \mathcal{L}_z = \begin{pmatrix} 0 & -1 & 0 \\ 1 & 0 & 0 \\ 0 & 0 & 0 \end{pmatrix} \, .
\end{equation}
Note that they can be defined in terms of the Levi-Civita symbol: $(\mathcal{L}_a)_{bc} = - \epsilon_{abc}$. The derivatives of $\mathbf{n} = \mathcal{R} \mathbf{n}'$ can then be expressed in terms of the covariant derivatives of $\mathbf{n}'$ as follows:
\begin{equation}
\mathcal{R}^{-1} \partial_\mu \mathbf{n} = (\partial_\mu + \mathcal{R}^{-1} \partial_\mu \mathcal{R}) \mathbf{n}' \equiv (\partial_\mu + \mathcal{A}^t_\mu) \mathbf{n}' \, ,
\end{equation}
where $\mu = 0$ denotes the temporal derivative and $\mu = 1, 2$, and $3$ denote the spatial derivatives with respect to the coordinates $x, y$, and $z$, respectively. The effects of the background spin texture on top of which a magnon lives are captured by the matrices $\mathcal{A}^t_\mu$, which is skew-symmetric due to the orthonormality of the rotation matrix $\mathcal{R}$.

Explicitly, we have the following transformation of each derivative term in the equations of motion by keeping the terms that include the derivative of the order parameter $\mathbf{n}'$ at least once:
\begin{eqnarray}
\partial_\mu \mathbf{n} &=& \mathcal{R} (\partial_\mu + \mathcal{A}^t_\mu) \mathbf{n}' \, , \\
\partial_\mu^2 \mathbf{n} &=& \mathcal{R} (\partial_\mu + \mathcal{A}^t_\mu)^2 \mathbf{n}'  \, .
\end{eqnarray}

The transformation of the DMI is as follows:
\begin{eqnarray}
\boldsymbol{\nabla} \times \mathbf{n} &=& \mathcal{R} \mathcal{R}^\tr (\boldsymbol{\nabla} \times \mathbf{n}) \\
&=& \mathcal{R} \left[\mathcal{R}^\tr (\boldsymbol{\nabla} \times \mathcal{R} \mathbf{n}') \right] \, .
\end{eqnarray}
The factor inside the square bracket is given by
\begin{eqnarray}
\left[\mathcal{R}^\tr (\boldsymbol{\nabla} \times \mathcal{R} \mathbf{n}') \right]_a &=& \mathcal{R}^\tr_{ac} \epsilon_{cde} \partial_d \mathcal{R}_{ef} n'_f \\
&=& \left[ \mathcal{A}^\text{d}_b \partial_b \mathbf{n}' \right]_a \, ,
\end{eqnarray}
where 
\begin{equation}
\mathcal{A}^d_b = \mathcal{R}^\tr \mathcal{L}_b \mathcal{R} \, .
\end{equation}

In terms of the $\mathbf{n}'$, the Landau-Lifshitz-Gilbert Eq.~(3) is given by
\begin{equation}
s (\partial_t + \mathcal{A}^t_0) \mathbf{n}' + \rho \mathbf{n}' \times (\partial_t + \mathcal{A}^t_0)^2 \mathbf{n}' = \mathbf{n}' \times A \left( \partial_i + \mathcal{A}^t_i - (D/A) \mathcal{A}^d_i \right)^2 \mathbf{n}' \, ,
\end{equation}
when we keep the terms that include the temporal or spatial derivative of $\mathbf{n}'$ at least once by focusing on high-energy magnons. For small deviations from the equilibrium, $\mathbf{n}' \approx \hat{\mathbf{z}} + n'_x \hat{\mathbf{x}} + n'_y \hat{\mathbf{y}}$ with $|n'_x|, |n'_y| \ll 1$, we obtain the following equation of motion for $n_+ \equiv n'_x - i n'_y$:
\begin{equation}
- i s (\partial_t + i \phi / \hbar) n_+ - \rho (\partial_t + i \phi / \hbar)^2 n_+ = - A \left[ \partial_i - i (a^t_i + a^d_i) / \hbar \right]^2 n_+ \, ,
\end{equation}
where the scalar potential $\phi$, the vector potential $\mathbf{a}^t$ from the spin texture, the vector potential $\mathbf{a}^d$ from the DMI are given by
\begin{equation}
\phi \equiv \hbar (\mathcal{A}^t_0)_{12} \, , \quad a^t_i \equiv - \hbar (\mathcal{A}^t_i)_{12} \, , \quad a^d_i \equiv (\hbar D/A) (\mathcal{A}^d_i)_{12} \, .
\end{equation}
One set of explicit expressions of them is given by $\phi = - \hbar \cos \theta_0 \partial_t \phi_0$, $\mathbf{a}^t = \hbar \cos \theta_0 \boldsymbol{\nabla} \phi_0$, and $\mathbf{a}^d = - (\hbar D/A) \mathbf{n}_0$.

\section{The dynamics of a nonrelativistic charged particle in the presence of a uniform magnetic field}

In the presence of a strong perpendicular magnetic field, the bands of a charged particle in two-dimensional systems form the Landau levels. Here, we summarize the known results for the Landau levels of a nonrelativistic charged particle in the presence of a uniform magnetic field~\cite{[][{, and references therein.}] Girvin1999}, which has been adopted for a ferromagnetic magnon previously~\cite{vanHoogdalemPRB2013}. The pertinent Schr{\"o}dinger equation is given by
\begin{equation}
\label{eq:s-eq}
i \hbar \partial_t \psi = \left[ \frac{1}{2 m} \left( \frac{\hbar}{i} \boldsymbol{\nabla} - q \mathbf{a} \right)^2 \right] \psi \, ,
\end{equation}
where $q = \pm 1$ parametrizes the charge of the particle. The external magnetic field $b_z = (\boldsymbol{\nabla} \times \mathbf{a}) \cdot \hat{\mathbf{z}}$ perpendicular to the plane is characterized by the magnetic length, which is given by $l = \sqrt{\hbar / |b_z|}$. The corresponding cyclotron frequency is given by $\omega_c = \hbar / m l^2$.

 The solution to Eq.~(\ref{eq:s-eq}) is then given by $\psi (\mathbf{r}, t) = \psi_{n n'} (\mathbf{r}) \exp(- i \epsilon_n t / \hbar)$, with the energy of the $n$th Landau level ($n = 0, 1, 2, \cdots$)
 \begin{equation}
 \epsilon_n = \hbar \omega_c (n + 1/2) = (\hbar |b_z| / 2m) (2 n +1) \, ,
 \end{equation}
and the spatial part $\psi_{n n'} (\mathbf{r})$ is the eigenfunction of the right-hand side of Eq.~(\ref{eq:s-eq}) with the eigenvalue $\epsilon_{n}$. See Ref.~\cite{Girvin1999} for the explicit construction of the spatial part $\psi_{n n'}$. The number of each Landau level's states, which are indexed by the integer $n'$ in $\psi_{n n'}(\mathbf{r})$, is given by the total number of magnetic flux quanta through the plane. The semiclassical equations of motion for the wavepacket localized at the position $\mathbf{r}$ and the momentum $\mathbf{k}$ are given by
\begin{eqnarray}
\dot{\mathbf{r}} &=& \frac{1}{\hbar} \frac{\partial \epsilon_n (\mathbf{k})}{\partial \mathbf{k}} - \dot{\mathbf{k}} \times \boldsymbol{\Omega}_n (\mathbf{k}) \, \\
\hbar \dot{\mathbf{k}} &=& - \boldsymbol{\nabla} U(\mathbf{r}) \, ,
\end{eqnarray}
where $\boldsymbol{\Omega}$ is the Berry curvature defined in terms of the magnetic Bloch wavefunctions~\cite{XiaoRMP2010, MurakamiJPSJ2017}. For flat Landau levels, the Berry curvature is uniform, and the Chern number, defined as the integral of the $z$-component of the Berry curvature, is determined by the product of the charge and the direction of the magnetic field: $\nu_n = \int_\text{BZ} d k_x d k_y \boldsymbol{\Omega}_n \cdot \hat{\mathbf{z}} / 2 \pi = - \sgn(q b_z)$. 
\end{widetext}

\end{document}